\documentstyle[12pt]{article}
\begin{document}

\def\eel#1{\label{#1}\end{equation}}
\newcommand{\be}{\begin{equation}}
\newcommand{\ba}{\begin{array}}
\newcommand{\ee}{\end{equation}}
\newcommand{\ea}{\end{array}}
\newcommand{\form}[1]{(\ref{#1})}
\newcommand{\med}{\frac{1}{2}}
\newcommand{\T}{T_{\rm bh}}
\newcommand{\M}{M_{\rm Pl}}
\newcommand{\Bu}{B_\uparrow}
\newcommand{\Bd}{B_\downarrow}
\newcommand{\bh}{black hole}

\begin{titlepage}
\title{\Large{\bf Quantum Black Holes}
\thanks{Work partially supported by MEC-FPI grant.}}

\author{
{\bf Juan Garc\'{\i}a--Bellido}\thanks{
 e--mail: bellido@slacvm.bitnet} \\ \normalsize
Theory Group, \ Physics Department\\ \normalsize
Stanford University \\ \normalsize
Stanford, CA 94305-4060 }

\date{}
\maketitle
\begin{abstract}
Following Mukhanov's pioneering work, we propose the hypothesis
that a \bh\ spontaneously emits radiation in quanta of
entropy of one {\em bit}. It then follows, using
standard thermodynamical arguments, that the black hole
should be quantized. We give the spectrum of the quantum
black hole, the degeneracy of the levels and their width.
There is a ground state of zero entropy, area $4\ln 2\
\M^{-2}$ and
mass $(\ln 2/4\pi)^{1/2} M_{\rm Pl}$. It may be a stable
particle, the end--point of Hawking
evaporation. Using the equivalence between \bh s and
white holes, we find that the Universe could have
started in a ground state of density $3/(8\ln 2)\ \M^4$.
\end{abstract}

\vskip-14.5cm
\rightline{\bf SU--ITP--93-4}
\rightline{\bf IEM--FT--68/93}
\rightline{\bf hep-th/9302127}
\rightline{\bf February 1993}
\vskip3in

\end{titlepage}

\newpage

According to information theory \cite{INF}, the entropy
or lack of information associated with the question ``does
a particle exist inside a black hole?" is given by
\be
S = - p \ln p - (1-p) \ln (1-p), \hspace{5mm}
0 \leq p \leq 1.
\eel{SIT}
If we knew the answer with absolute certainty, probability
$p=1$, then the entropy would be $S=0$. On the other hand, the
maximum ignorance (about the answer) corresponds to the maximum
of this function, {\em i.e.} equal probabilities
$p=1/2$ and thus $S=\ln 2$, one {\em bit} of information.

Hawking's ``Principle of Maximum Ignorance" \cite{PRD}
suggests that a \bh's event horizon acts as an information
blocking mechanism which forces one to assign equal probabilities
to the question above \cite{FN1}. Therefore, following
the pioneering work of Mukhanov \cite{MUK}, we assume
that the radiation emitted {\em spontaneously} by a \bh\
carries a quantum of entropy, a bit. It then follows,
using standard thermodynamical arguments, that
the \bh\ itself must be quantized.

The radiation modes are thus characterized
by a frequency $\omega$, a charge $q$, an azimuthal quantum
number $m$ and a quantum of entropy, a {\em bit}, satisfying
Bekenstein's \cite{BEK} first law of black hole thermodynamics
(in units $c=\hbar=k=G=1$),
\be
\omega - q\Phi - m\Omega  = \T\  \ln 2,
\eel{1LT}
where $\Phi$, $\Omega$ and $\T$ are the electrical potential,
the rotational frequency and the temperature of the black hole,
respectively.

The mean number of quanta emitted by a black hole in a given
mode is given by Hawking's well known result \cite{HAW}
\be
<n> = \frac{\Gamma}{e^x - 1},
\eel{HAW}
where $\Gamma$ is the "absorptivity" of the black hole and
\be
x = \frac{\omega - q\Phi - m\Omega}{T_{\rm bh}}.
\eel{XEQ}
Eq.\form{HAW} is just what would be emitted by a {\em grey}
body whose $\Gamma$ for a given mode coincides with that of
the black hole, and which has a well defined temperature $\T$.

Bekenstein and Meisels \cite{BKS} are then able, using very
general thermodynamical arguments, to derive the expression
for the Einstein $A$ and $B$ coefficients for the black hole
in terms of $\Gamma$ and $x$,
\be\ba{c}
{\displaystyle
A_\downarrow = B_\downarrow = \frac{\Gamma}{e^x - 1} },\\[2mm]
{\displaystyle
B_\uparrow = \frac{e^x \Gamma}{e^x - 1} },
\ea\eel{ABG}
and the related identities,
\be\ba{c}
{\displaystyle
\Gamma = B_\uparrow - B_\downarrow },\\[2mm]
{\displaystyle
\frac{B_\uparrow}{B_\downarrow} = e^x },
\ea\eel{GBB}
where $B_\uparrow$ and $B_\downarrow$ are the induced
absorption and emission coefficients, and $A_\downarrow$ the
spontaneous emission coefficient of the \bh. We will
assume that these coefficients \form{ABG} have the usual
statistical interpretation in terms of internal \bh\
configurations \cite{BKS}, and that the quantum mode
\form{1LT} is the result of transitions between
contiguous internal levels of the \bh.
In that case we have
\be
\frac{\Bu}{\Bd} = 2.
\eel{BB2}
We know from statistical physics that the ratio of the
Einstein $B$ coefficients is exactly the ratio of the
degeneracy of the respective levels, which we will denote
by $W_n$ and $W_{n-1}$. Thus we deduce
\be
W_n = 2\ W_{n-1}.
\eel{WW2}
This relation must be valid for {\em all} internal levels $n$,
since the arguments used to deduce it are independent of
$n$. Then we may write $W_n = c\ 2^n$, where $c$ is a constant.
We can now define the entropy of a given level, which is
related to the number of independent internal configurations
of that level, as the usual Boltzmann entropy
\be
S_n = \ln W_n = n \ln 2 + \ln c.
\eel{SLN}
We may fix the constant $c$ by requiring that the ground
level ($n=1$) has zero entropy, or equivalently, is
non--degenerate,
\be\ba{c}
{\displaystyle
S_n = (n - 1) \ln 2 },\\[2mm]
{\displaystyle
W_n = 2^{n-1} },
\ea\eel{SWN}
strongly suggesting that the entropy of a black hole
is quantized, and that each internal level has degeneracy
$W_n$, precisely the number of integer partitions
of $n$, as would be expected if we had {\em no} information,
due to the existence of an event horizon,
of the internal configurations that constitute each level.
These relations were first obtained by Mukhanov \cite{MUK},
using the correspondence principle, in the limit $n\gg 1$.
See also \cite{LNC}.

We know \cite{BEK,HAW} that the surface area, and therefore the
mass, of the black hole are related to its entropy by
very simple formulas (We will consider only Schwarzschild
\bh s from now on). Therefore, the entropy quantization
\form{SWN} ensures also an area and mass quantization \cite{FN2},
\be\ba{c}
{\displaystyle
A_n = n \ 4 \ln 2, }\\[2mm]
{\displaystyle
M_n = \left(\frac{\ln 2}{4\pi}\right)^{1/2} \sqrt{n}.}
\ea\eel{AMN}
Note that the minimal possible increase of the black hole
area is $4\ln 2$, a result found by Bekenstein \cite{BEK}.
Therefore, a black hole of mass $M$ may consist of infinitely
many quantum internal states or configurations, disposed
in levels of definite area, mass and entropy, which are
exponentially degenerate \form{SWN}.

One may ask
\cite{MUK} what are the lifetimes $\tau_n$ of these levels,
or equivalently, their quantum widths $\Gamma_n$.
These can be computed using again thermodynamical
arguments. We know that a hot body of averaged absorptivity
$\Gamma$ and area $A$ at a temperature $T$
emits energy at a rate given by Stefan--Boltzmann law,
$dE/dt = \sigma \Gamma A T^4$, where $\sigma$ is the
Stefan--Boltzmann constant, which in our units reads
$\sigma = \pi^2/60$. The energy emitted by a black hole
in the form of Hawking radiation thus corresponds to a
decrease in its mass
\be
\frac{d M}{d t} = - \frac{\Gamma}{15360\pi}\
\frac{1}{M^2},
\eel{DMT}
where we have substituted $A_{\rm bh} = 16 \pi M^2$
and $\T\ = 1/(8\pi M)$. Note that it gives the expected
Hawking evaporation rate \cite{PAG}. We now assume that
each level $n$ emits radiation as if it were a hot
body of area $A_n$ at temperature $T_n$, and whose
mass--loss rate is dominated by $dM_n/dt \simeq -
\omega_{\rm max}/\tau_n = - 2.82 \ T_n\ \Gamma_n$,
using Wien's displacement law. From \form{DMT} we find
\be
\Gamma_n \simeq \frac{\Gamma}{1920}\ \frac{1}{3M_n} =
\frac{8\pi}{3\ln 2}\ \frac{\Gamma}{1920}\ \Delta
M_{n,n-1} \ll \Delta M_{n,n-1}
\eel{GMN}
where we have used Eq.\form{1LT},
$\Delta M_{n,n-1}\equiv \omega_{n,n-1} =
\ln 2/(8\pi M_n)$. We have thus proven, for all $n$, that
the width of the internal levels of the black hole are
much smaller than the spacing between levels \cite{MUK}.
We can therefore speak of a true quantum black hole.

We now propose \cite{MUK} that Hawking radiation is simply
the averaged radiation emitted by a black hole, composed
of infinitely many discrete transitions (very similar to
the radiation emitted by an atom). Unfortunately we do not
have the experimentally observed spectral lines of a
black hole to compare with, since the energy resolution
of the observed gamma ray bursts is insufficient to resolve
the black hole spectrum \cite{GRB}. However, there is a
fundamental difference with Hawking's spectrum, since we can
deduce from \form{SWN} and \form{AMN} that there exists a
non--degenerate ground state $(n=1)$, with
minimum area $A = 4\ln 2\ \M^{-2}$,
{\em zero} entropy, and mass
$M = (\ln 2/4\pi)^{1/2} M_{\rm Pl}$. All other states are
unstable (Hawking radiation is the result of their
transitions to the ground state). Of course, we cannot
ensure the stability of the ground state under quantum
gravity effects but, as far as the above description is
concerned, such a state is stable. Thus, the end--point
of Hawking's evaporation may not be a naked singularity
but a fundamental particle \cite{PLB}.

Furthermore, as pointed out in Ref.\cite{BKS}, the
relation between the B coefficients in Eq.\form{GBB}
is a direct consequence of time reversibility of
the \bh\ emission and absorption, supported by
Hawking's view \cite{HWK} that a \bh\ and a white hole
are equivalent. Therefore, our
former quantum description of a black hole could
just as well apply to a white hole, like the Big Bang.
This would mean that the Universe could have started
in a ground state of Planck density, and evolved towards
our present Universe through emission of quantum
radiation. We can compute the energy density of this
ground state by assigning to a de Sitter Universe a
temperature $T_{\rm dS} = H/2\pi$ \cite{GIB}, where
H is the Hubble parameter, $H^2 = 8\pi\rho/3$ in
Planck units. Comparing the area of the de Sitter horizon,
$A_{\rm dS} = 4\pi H^{-2}$, with the minimum quantum area,
see Eq.\form{AMN}, we find that the energy density is
quantized
\be
\rho_n = \frac{3}{8\ln 2}\ \frac{1}{n}\ \M^4.
\eel{DGS}
So the Universe could have started in a ground state very
close to Planck density, and have evolved until today,
where $n\gg 1$ and the cosmological constant is small
$\Lambda \sim 10^{-120} \M^4$  \cite{SCH}.

Of course, these results, obtained
using standard thermodynamical arguments, should be
confirmed by formal quantum field--theoretic
calculations. Note that we still do {\em not} have
a quantum theory of gravity nor a quantum mechanics
of black holes, but at least we know that there is a
deep connection between gravitation, quantum theory
and thermodynamics.

\section*{Acknowledgements}

The author thanks
L. Susskind, A. Linde and V. Mukhanov for various
enlightening discussions. He is also greatly endebted
to J. Bekenstein for useful comments on the role of
information theory in \bh\ quantization.


\begin{thebibliography}{99}
\bibitem{FN2} Note that we are assuming $S=A/4 - \ln 2$, which
gives no contradiction between the second law of thermodynamics
and Hawking's area theorem \cite{PRL}, since $S \geq S_1 + S_2$
implies $A \geq A_1 + A_2 - 4\ln 2$. If we take initial
minimum \bh\ areas, $A_i = 4\ln 2$, then the final area
satisfies $A \geq A_{\rm min} = 4\ln 2$, consistently.
\end{thebibliography}
\end{document}